\documentclass[a4paper, 11pt]{article}

\usepackage{graphicx}
\usepackage{xcolor}
\usepackage{slashed}
\usepackage{ amssymb }
\usepackage{mathtools}
\usepackage{latexsym}
\usepackage{textgreek}
\usepackage{verbatim}
 \usepackage[utf8]{inputenc}
 \usepackage{amsmath}
\usepackage{amsfonts}
\usepackage{verbatim}
\usepackage{graphicx}
\usepackage{subfigure}
\usepackage{amssymb}
\usepackage[T1]{fontenc}
\usepackage{slashed}
\usepackage{color}
\usepackage[numbers,sort&compress]{natbib}

\def\beq{\begin{eqnarray}}
\def\eeq{\end{eqnarray}}

\def\({\left(}
\def\){\right)}

\newcommand{\be}{\begin{equation}}
\newcommand{\ee}{\end{equation}}
\newcommand{\la}{\langle}
\newcommand{\ra}{\rangle}
\def\ea{\end{eqnarray}}
\def\ba{\begin{eqnarray}}

\def\beq{\begin{eqnarray}}
\def\eeq{\end{eqnarray}}

\def\({\left(}
\def\){\right)}

\def\p{\partial}
\def\la{\langle}
\def\ra{\rangle}

\def\lsim{\mathrel{\rlap{\lower3pt\hbox{\hskip0pt$\sim$}}
     \raise1pt\hbox{$<$}}} 
\def\gsim{\mathrel{\rlap{\lower4pt\hbox{\hskip1pt$\sim$}}
     \raise1pt\hbox{$>$}}}

\def\lsim{\mathrel{\rlap{\lower3pt\hbox{\hskip0pt$\sim$}}
     \raise1pt\hbox{$<$}}} 
\def\gsim{\mathrel{\rlap{\lower4pt\hbox{\hskip1pt$\sim$}}
     \raise1pt\hbox{$>$}}}

\begin{document}

\begin{center}{\Large \bf{Coherent States in Gauge Theories: \\ Topological Defects and Other Classical Configurations
}}

 \vspace{1truecm}
\thispagestyle{empty} \centerline{\large  {Lasha  Berezhiani $^{1,2}$, Gia Dvali $^{1,2}$ and Otari Sakhelashvili $^3$}
}

 \textit{$^1$Max-Planck-Institut f\"ur Physik, Boltzmannstra{\ss}e.~8,
85748 Garching, Germany\\
 \vskip 5pt
$^2$Arnold Sommerfeld Center, Ludwig-Maximilians-Universit\"at, \\Theresienstra{\ss}e 37, 80333 M\"unchen, Germany\\
 \vskip 5pt
 $^3$Sydney Consortium for Particle Physics and Cosmology, \\
School of Physics, The University of Sydney, NSW 2006, Australia
 }

\end{center}  
 
\begin{abstract}

We present a formulation of coherent states as of consistent quantum description of classical configurations in the BRST-invariant quantization of electrodynamics. 
The quantization with proper gauge-fixing is performed on the vacuum of the theory, whereas other backgrounds are obtained as BRST-invariant coherent states. One of the key insights is the possibility of constructing the coherent states of pure-gauge configurations. This provides a coherent state understanding of topologically non-trivial configurations in gauge theories, and makes number of features, such as the suppression of transitions between topologically-distinct sectors, very transparent at full quantum level. As an example, we construct the Nielsen-Olesen string as a BRST-invariant coherent state. The  Abelian pure-gauge configurations can also be viewed as useful analogs for a set of space-times related by coordinate reparameterizations in General Relativity.

\end{abstract}

\newpage
\setcounter{page}{1}

\renewcommand{\thefootnote}{\arabic{footnote}}
\setcounter{footnote}{0}

\linespread{1.1}
\parskip 4pt

\section{Introduction} 

One of the most important tasks of quantum field theory  
is to achieve a full microscopic description of 
would-be classical field configurations. 
 This is especially important for non-linear 
 systems in strong field regimes such as black holes,
cosmological space-times and various solitons.  

There is an ongoing research program of a quantum corpuscular resolution of these systems in terms of quantum coherent states 
 (or condensates) of high occupation number of 
  quanta \cite{Dvali:2011aa, Dvali:2012en,Dvali:2013eja,  Dvali:2014gua, Dvali:2015jxa,Berezhiani:2016grw, 
 Dvali:2017eba, Dvali:2020wqi,Berezhiani:2020pbv,Berezhiani:2021gph,Berezhiani:2021zst, Dvali:2022vzz,Berezhiani:2023uwt, Berezhiani:2024boz}.   
This resolution is essential for 
on one hand reproducing the (semi-)classical dynamics as an emergent description of quantum systems and on the other hand for calculating the departures from this effective description. 
 While the coherent state description of classical fields for 
 which nonlinearities have negligible impact has a long history \cite{Glauber:1963tx,Sudarshan:1963ts,Kibble:1965zza} (for the early review see \cite{Zhang:1990fy,Zhang:1999is}), 
   the  systems with strong collective couplings reveal fundamentally new features.  
   
  In particular, the studies in 
  \cite{Dvali:2013eja, Dvali:2014gua, Berezhiani:2016grw, 
 Dvali:2017eba, Dvali:2018xpy, Dvali:2020wft}
  have highlighted that such 
  systems are subjected to a phenomenon
 of ``quantum breaking'', a complete break-down of the semi-classical picture after a certain critical time, called the ``quantum break-time'' \cite{Dvali:2013vxa}.
   This departure has important implications for cosmology as 
   well as for the black hole physics. 
   In particular, in cosmology the quantum breaking invalidates de Sitter as a valid vacuum, whereas for a black hole it leads to a breakdown of 
   Hawking's semi-classical regime. 
    
 Moreover, in systems of high capacity of information storage, such as 
  black holes or a de Sitter like universe,  
  the coherent state picture reveals a specific mechanism of 
  quantum breaking in the form of a so-called ``memory-burden'' effect 
  \cite{Dvali:2018xpy, Dvali:2020wft}:  a slow-down of the system's decay due to a  stabilizing back-reaction from the carried quantum information.  

 At the same time, the coherent state resolution  has important 
 implications for non-perturbative objects in gauge theories. 
 In particular, it was shown in \cite{Dvali:2015jxa} that the topological charge acquires 
 an intrinsically quantum meaning in terms of a singular occupation number of the infrared (zero-momentum) modes.   This fact gives 
a transparent meaning to the suppression of quantum transitions 
between distinct topological sectors.

Furthermore,  as shown in \cite{Dvali:2019jjw, Dvali:2019ulr, Dvali:2020wqi}, the corpuscular resolution of solitons reveals 
the universality of the upper bound on the microstate entropy imposed 
by unitarity,  which has the area-form strikingly similar to 
the Bekenstein-Hawking entropy of a black hole. 

 The corpuscular picture was generalized to Euclidean configurations such as instantons in  \cite{Dvali:2019ulr}. Despite the fact that instantons are processes rather than the states, their corpuscular 
 resolution has a very well defined physical meaning and 
 is in one-to-one correspondence  with the corpuscular resolution of 
 the gauge configurations that it connects via tunnelling.   
   
 The resolution of strong field backgrounds as of multi-particle coherent 
     states allows to understand the would be non-perturbative processes 
      in terms of multi-particle scattering amplitudes that are fully perturbative in four-point quantum coupling $\alpha$.  For example, 
      this has been demonstrated both for Hawking radiation of black  
  holes \cite{Dvali:2011aa}  and for Gibbons-Hawking radiation in de Sitter space \cite{Dvali:2013eja}.  Ordinarily, these processes are 
  considered as non-perturbative semi-classical processes. 
 However, the resolution of the strong classical 
  gravitational field  in terms of a coherent state of gravitons 
  reveals their quantum nature as multi-graviton scattering processes
   that are perturbative in gravitational coupling $\alpha_{\rm gr}$.  
   In particular it was shown \cite{Dvali:2017eba} that the Boltzmann thermal suppression factor  ${\rm exp}(-m/T)$ for the production 
  of particles much heavier than the corresponding  Hawking
  or Gibbons-Hawking temperatures can be understood as the 
  suppression of multi-graviton scattering processes that are exponential 
  in inverse gravitational coupling ${\rm exp}(-\frac{1}{\alpha_{\rm gr}})$. 
    Such exponential suppression of many-to-few graviton processes
has been derived by explicit computations in 
\cite{Dvali:2014ila, Addazi:2016ksu}.

   Analogously, for non-gravitational theories,  it was shown 
 \cite{Dvali:2022vzz}  
   that the non-perturbative processes that are suppressed by ${\rm exp}(-\frac{1}{\alpha})$ in reality 
   represent the multi-particle scattering processes with 
   the number of participating quanta $N = 1/\alpha$. 
   This  was explicitly demonstrated  for particle creation in the oscillating field as well as for Schwinger pair creation. It was shown that
   multi-particle processes, perturbative in $\alpha$, recover known 
   results, such as Schwinger rate in the semi-classical limit. 
  
     Understanding the particle-creation by classical strong-field 
   backgrounds in terms of multi-particle  amplitudes  of coherent states allows to capture the corrections to semi-classicality that lead to back-reaction and to the subsequent quantum breaking.

In this direction, recently a significant progress has been made in understanding 
various aspects of the consistent construction and dynamics of coherent states within interacting field theories \cite{Berezhiani:2020pbv,Berezhiani:2021gph,Berezhiani:2021zst,Berezhiani:2023uwt}. 

The coherent state resolution of non-perturbative backgrounds 
 adopted in \cite{Dvali:2011aa, Dvali:2012en, Dvali:2013eja,  Dvali:2014gua, Berezhiani:2016grw, 
 Dvali:2017eba, Berezhiani:2021zst, Berezhiani:2024boz}  rests on two key ingredients.  
   The idea is to quantize the system on a unique vacuum of Minkowski and treat other classical backgrounds as 
   the coherent states constructed on top of this vacuum. 
   In gravity, this view, is independently supported by the 
   $S$-matrix reasoning which singles out the Minkowski space as 
  the valid $S$-matrix vacuum \cite{Dvali:2020etd}.  

   The consistent implementation of both ingredients was carried out in  our two previous papers  \cite{Berezhiani:2021zst, Berezhiani:2024boz}
 within the BRST-invariant framework.  
The present work is a continuation of this line. 
Namely, we follow-up on \cite{Berezhiani:2021zst}, where we studied the aforementioned subject within Abelian gauge theories, namely in quantum electrodynamics (QED) with scalar charges and linearized gravity. It was demonstrated there that the physical states must come with the entourage of appropriate dressing fields by consistency. The analysis was carried out within the so-called BRST-invariant quantization \cite{Kugo:1979gm}, which is an equally robust framework for the $S$-matrix and the real-time correlation function analysis within the background field method; not only for QED but for non-Abelian gauge theories and General Relativity as well.

One of the important extensions of the previous work provided here, is the discussion of coherent states of pure-gauge configurations. This is the essential step towards a better understanding of coordinate reparameterization in quantized General Relativity, as discussed in \cite{Berezhiani:2024boz}. The point is that, even though the gauge is fixed upon quantization via Faddeev-Popov procedure, we are nevertheless free to construct coherent states of pure-gauge configurations consistent with the physicality condition. Although the states that differ by such configurations have identical $S$-matrix elements, the expectation value of the gauge field (as well as any other non-gauge-invariant correlation function) is not. We show this to be a straightforward consequence of the adopted formalism.

The reason these are important points, especially for the theory of gravity, stems from our philosophy of thinking of the quantum state of the system in its entirety. In other words, we quantize the theory before positing the background as a quantum state. This way of reasoning selects a certain gauge from the beginning, in which the canonical quantisation was performed. Therefore, it sets the gauge of the background and perturbations in a connected way. There is still a freedom to readjust the gauge for perturbations as needed, e.g. in the path-integral, but only for the computation of gauge-invariant quantities \cite{Weinberg}. However, this could have important implications for computing loop corrections to cosmological correlation functions within the fundamental framework \cite{Berezhiani:2024boz} similar to the one adopted in this work, due to the fact that the gauge-invariant quantities usually invoked for such computations are gauge-invariant only to the linear order in perturbations. The caveats of working with gauge non-invariant variables has been previously discussed in \cite{Garriga:2007zk} in the context of computing the quantum back-reaction on cosmological backgrounds. In this work, we demonstrate the relevance of such considerations within quantum electrodynamics.

Furthermore, we perform the explicit construction of topological defects as BRST-invariant coherent states, within an Abelian gauge theory. 
In this context, we explicitly verify the point made in \cite{Dvali:2015jxa}
that the topological charge in the quantum picture is accounted 
for by the singularity of the occupation number of zero-momentum
modes. We show that despite the mass gap, due to the gauge redundancy, the infinite occupation number is fully  consistent 
with the finite energy of the soliton. 

Our construction can also be used for a  BRST-invariant realization 
of the proposal of  \cite{Dvali:2019ulr} for  
corpuscular resolution of an instanton, since the Nielsen-Olesen
vortex in $2+1$-dimensions is mapped on an instanton of a $1+1$-dimensional Higgs theory. 

 Before proceeding, we wish to cite the earlier construction of topological defects as coherent states performed in \cite{Taylor:1977bq}, as well as 
 more recent work \cite{Evslin:2023qbv,Evslin:2024czr,Ogundipe:2024chv,Evslin:2024zkx}.

The paper is organized as follows. Sec. \ref{qed_sec} overviews some of the essential aspects of the BRST-invariant formulation of quantum electrodynamics. Sec. \ref{class_source} performs the explicit time-evolution of a coherent state sourced by the fundamentally classical source, albeit in pure QED without dynamical charges. Sec. \ref{scalar_qed} introduces dynamical charges in the form of a complex scalar field and discusses the consistent construction of coherent states for such a matter. Sec. \ref{class_lim} demonstrates that there is a double-scaling classical limit in which the dressed coherent state of the scalar field factorizes into a direct product of a coherent state of the scalar and electromagnetic fields. Sec. \ref{gauge_fix} discusses implications of the gauge-fixing condition within the background field method and the construction of pure-gauge field configurations as BRST-invariant coherent states. Sec. \ref{top_def} is devoted to the construction of the Nielsen-Olesen string as a coherent state. In Sec. \ref{occup_numb}, we discuss the interpretation of the topological charge in terms of the occupation number. In Sec. \ref{instant} we generalize the discussion to instantons. We conclude in Sec. \ref{outlook}.

\section{BRST-invariant QED}
\label{qed_sec}

In quantum electrodynamics the BRST-quantization proceeds with the following gauge-fixed Lagrangian
\beq
\label{enmlag}
\mathcal{L}=-\frac14 \hat{F}_{\mu\nu}^2-\partial_\mu \hat{B}\hat{A}^\mu+\frac12\xi \hat{B}^2
+\partial_\mu \hat{\bar c}\partial^\mu \hat{c}-\hat{A}_\mu\hat{J}^\mu\,,
\eeq
with $\hat{J}^\mu$ standing for the matter current. This formulation enjoys the residual global symmetry with a Grassmannian parameter; see e.g. \cite{Kugo:1979gm}. The corresponding fermionic charge operator $\hat{Q}$ provides the following constraint defining the physical Hilbert space
\beq
\hat{Q}|f\ra=0\,.
\eeq
The Hamilton equations readily yield
\beq
&&\partial^\mu \hat{F}_{\mu\nu}=\partial_\nu \hat{B}+\hat{J}_\nu\,,\nonumber\\
&&\partial_\mu \hat{A}^\mu=-\xi \hat{B}\,.
\eeq
As it has been pointed out in \cite{Kugo:1979gm}, the matrix elements of these equations between physical states reduce to
\beq
&&\partial^\mu \la f_1|\hat{F}_{\mu\nu}|f_2\ra=\la f_1|\hat{J}_\nu| f_2\ra\,,\nonumber\\
&&\partial_\mu \la f_1 |\hat{A}^\mu| f_2\ra=0\,.
\eeq
This observation stems from the transformation property of the anti-ghost field
\beq
\label{bqc}
\hat{B}=i\{\hat{Q},\hat{\bar{c}}\}\,,
\eeq
under the BRST charge $\hat{Q}$. Namely, the expectation value of the gauge field in a physical state must satisfy the Lorentz condition for the gauge-fixing term introduced in \eqref{enmlag}. Following \cite{Berezhiani:2021zst}, the BRST-invariant coherent state can be built around the BRST-invariant vacuum $|\Omega\ra$ as
\beq
|\Psi\ra=e^{-i\int d^3 x\left(A_j^c \hat{E}_j-E^c_j\hat{A}_j+A_0^c\hat{B}\right)}|\Omega\ra\,.
\label{gaugestate}
\eeq
Here, quantities carrying the label 'c' are c-number functions that define the initial expectation values of the corresponding operators. In this coherent state, the expectation value of $\hat{B}$ vanishes, as it should due to the above-mentioned connection with the anti-commutator of the BRST charge and the anti-ghost field. This state is physical, i.e. BRST-invariant, as long as $\partial_jE_j^c=0$. It must be noted that this is an exact statement and does not assume the absence of charges.

\section{Classical Source}

\label{class_source}

Next, we would like to revisit the cosmological constant 'analog', considered in \cite{Berezhiani:2021zst}. In particular, we would like to analyze the evolution of the coherent state sourced by the fundamentally classical source. In \cite{Berezhiani:2021zst}, we discussed the particle pair-creation in such a state. Here, on the other hand, we would like to show the evolution of the coherent state in the absence of quantum charges, i.e. within pure electrodynamics with a homogeneous classical source. As we have shown in the previous work, this setup shares some of the important properties of the construction of de Sitter spacetime as a coherent state.

 We note that the corpuscular resolution of an electromagnetic field 
 created by a classical source in terms multi-photon state has been discussed previously 
 in \cite{Dvali:2011aa}, with explicit construction of corresponding coherent state performed in \cite{Muck:2013orm,Muck:2015dea}.

  Here we shall study such states within BRST-invariant construction of \cite{Berezhiani:2021zst}.  
In order to underline some of the caveats that emerge in gauge theories, we begin with the non-interacting scalar field prototype with the Hamiltonian
\beq
\hat{H}=\hat{H}_0+\int d^3 x \Lambda \hat{\Phi}\,,\qquad {\rm where}\qquad \hat{H}_0\equiv\int d^3 x \left( \frac{1}{2}\hat{\Pi}^2+\frac{1}{2}(\partial_j\hat{\Phi})^2 \right)\,.
\eeq
In other words, we have a massless scalar field with a homogeneous classical source ($\Lambda=const$) at hand. Classically, the initial homogeneous field displacement $\Phi_0$ with a vanishing initial momentum $\Pi_0=0$ will roll as
\beq
\Phi_{\rm cl}=\Phi_0-\frac{1}{2}\Lambda t^2\,.
\label{clPhi}
\eeq
Following \cite{Berezhiani:2021zst}, we construct coherent states associated with such configurations over the vacuum of $\hat{H}_0$, satisfying
\beq
\hat{H}_0|0\ra=0\,.
\eeq
Namely, in the Schr\"odinger picture, we have
\beq
|C\ra (t=0)=e^{-i\Phi_0\int d^3 x \hat{\Pi}}|0\ra\,.
\eeq
In the non-interacting theory at hand the exact Hamiltonian flow of this state is straightforward to obtain. Computation of few commutators leads us to
\beq
|C\ra(t)=e^{-i\hat{H}t}\cdot e^{-i\Phi_0\int d^3 x \hat{\Pi}}|0\ra=e^{-\frac{i}{2}\int_{t'=0}^{t'=t} d^4 x \Lambda \Phi_{\rm cl}(t')}\cdot e^{-i\int d^3x\left( \Phi_{\rm cl}\hat{\Pi}-\dot{\Phi}_{\rm cl}\hat{\Phi} \right)}|0\ra\,,
\label{linearscalar}
\eeq
where $\Phi_{\rm cl}$ is given by \eqref{clPhi}. In other words, the only additional feature compared to what we could have naively guessed is the additional phase (the first exponential). Interestingly, it represents the classical action due to the linear potential.

Notice that we might as well have taken $|0\ra$ as a consistent initial state. This freedom is absent in gauge theories, in fact as we have shown in \cite{Berezhiani:2021zst} the physicality constraint enforces the presence of the coherent gauge field to dress the classical source.

In order to repeat the same exercise for the simplest gauge theory, we consider the following Hamiltonian
\beq
\hat{H}=\hat{H}_0+\int d^3 x \rho \hat{A}_0\,, \quad {\rm where} \quad \hat{H}_0\equiv \int d^3x \left(\frac{1}{2}\hat{E}_j^2+\frac{1}{4}\hat{F}_{ij}^2-\hat{A}_j\partial_j \hat{B}-\frac{\xi}{2}\hat{B}^2+\hat{E}_j\partial_j \hat{A}_0 \right)\,
\eeq
and $\rho=const$.
Here $\xi$ is the gauge-fixing parameter and we have dropped the Faddeev-Popov ghosts as they decouple completely for the theory in question; see e.g. \cite{Berezhiani:2021zst} for the Lagrangian formulation, discussion of symmetries and canonical variables.

The BRST charge operator for $\hat{H}$ differs from the one for $\hat{H}_0$. As such, the vacuum of $\hat{H}_0$ is no longer annihilated by the BRST charge operator in the presence of the classical source. Nevertheless, the physical coherent state can be constructed over the said unphysical vacuum as
\beq
|\rho\ra=e^{i\int d^3 x E_j^c\hat{A}_j}|0\ra\,, \quad {\rm with}\qquad \hat{H}_0|0\ra=0\, \quad {\rm and} \quad E_j^c=\frac{\rho}{3}x_j\,.
\eeq
Similar to the case of the scalar field, discussed above, the exact time evolution readily follows
\beq
|\rho\ra(t)=e^{-i\hat{H}t}|\rho\ra=e^{-i\xi\int d^3x \frac{t^2}{2}\rho^2}\cdot e^{i\int d^3x \left( E_j^c \hat{A}_j-t E_j^c \hat{E}_j+\frac{t^2}{2}E_j^c\partial_j\hat{B}-t\partial_j(x_j \hat{A}_0) \right)}|0\ra\,.
\eeq
Notice the appearance of the phase factor which, similar to \eqref{linearscalar}, comes with an infinite volume factor for the homogeneous charges. The implications of the nontrivial part of the state are straightforward to obtain and give rise to the classical evolution of the vector potential, i.e.
\beq
\la \hat{E}_j \ra = \frac{1}{3}\rho x_j\,,\qquad \la \hat{A}_j\ra=\frac{1}{3}\rho x_j t\,, \qquad \la \hat{A}_0 \ra=\frac{1}{2}\rho t^2\,.
\eeq
Not surprisingly, this is precisely the behaviour one obtains by simply solving classical Maxwell's equations.

\section{Quantum Scalar Source}
\label{scalar_qed}

The Hamiltonian of QED with dynamical scalar charges is as follows
\beq
\label{hamilton}
\hat{H}=\int d^3 x\left[ \frac{1}{2}\hat{E}_j^2+\frac{1}{4}\hat{F}_{ij}^2+|\hat{\Pi}|^2+|D_j\hat{\Phi}|^2+V(|\hat{\Phi}|^2)\right.
-\p_j\hat{B}\hat{A}_j-\frac{\xi}{2}\hat{B}^2+\p_j\hat{A}_0 \hat{E}_j\nonumber\\
+ ig\hat{A}_0\left( \hat{\Phi}\hat{\Pi}-\hat{\Phi}^\dagger\hat{\Pi}^\dagger \right)+\hat{\Pi}_c\hat{\Pi}_{\bar c}+\p_j \hat{\bar{c}}~\p_j\hat{c}\Big]\,,~
\eeq
where the potential for the scalar field can be chosen at will, depending on the question at hand. If we are interested merely in massive scalar charges without further interactions then we can select $V=m^2|\Phi|^2$.

In order to give the background to the scalar field in the form of a BRST-invariant coherent states, we follow \cite{Berezhiani:2021zst} by utilizing Dirac operators
\beq
\label{diracphi}
&&\hat{\Phi}_{\rm g}\equiv\hat{\Phi}\cdot{\rm exp}\left(-ig\frac{1}{\nabla^2}\partial_j \hat{A}_j\right)\,,\\
&&\hat{\Pi}_{\rm g}\equiv\hat{\Pi}\cdot{\rm exp}\left(+ig\frac{1}{\nabla^2}\partial_j \hat{A}_j\right)\,.
\label{diracpi}
\eeq
Due to their gauge/BRST invariance, any state constructed using them would automatically satisfy the physicality condition of the BRST quantization. In particular, the scalar field coherent states take the following form
\beq
|C_{\rm g}\ra=e^{-i\int d^3 x\left( \Phi_{\rm c}\hat{\Pi}_{\rm g}-\Pi_{\rm c}\hat{\Phi}_{\rm g}+h.c. \right)}|\Omega\ra\,.
\label{dressedC}
\eeq
The invariance of Dirac's operators ensures that
\beq
\hat{Q} |C\ra_{\rm g}=0\,.
\eeq
Furthermore, such matter states automatically come dressed with an appropriate gauge field. In other words, if the c-number functions are chosen in such a way that they correspond to configurations with nonvanishing $U(1)$ charge then the expectation value of the electric field is exactly what one needs to satisfy the Gauss law.

We would like to finish this section by pointing out that the dressed coherent states \eqref{dressedC} have the following peculiar property. The c-number functions set the initial expectation values of gauge-invariant quantities, e.g. \eqref{diracphi}. However, the expectation value of the original field $\hat{\Phi}$ receives diverging quantum corrections on top of the expected classical background. Namely, we have
\beq
\la C _{\rm g}|\hat{\Phi}_{\rm g}| C_{\rm g}\ra=\Phi_{\rm c}\,, \qquad \Longleftrightarrow \qquad \la C _{\rm g}|\hat{\Phi}| C_{\rm g}\ra=\Phi_{\rm c}\cdot\la \Omega| e^{-ig\frac{1}{\nabla^2}\partial_j \hat{A}_j} |\Omega \ra = \Phi_{\rm c}+\mathcal{O}(g^2)\,,
\eeq
where the leading contribution to $\mathcal{O}(g^2)$ originates from the nonvanishing photon correlator at coincidence. These are quantum corrections, as they would be $\mathcal{O}(\hbar)$ if we were to reintroduce the Planck constant explicitly. Therefore, even at the initial time, $\la C _{\rm g}|\hat{\Phi}| C_{\rm g}\ra$ matches the classical background only in the classical limit and deviates by infinity away from it. This diverging departure must be combined with the field-renormalization, which should yield a finite result.

\section{Classical Limit}
\label{class_lim}

As it was demonstrated in \cite{Berezhiani:2021zst}, the naive expectation that the coherent state of scalar charges sources the coherent state of the corresponding electromagnetic field, in the form of the direct product of the two, is rejected by the physicality condition. In other words, it is straightforward to show that
\beq
\hat{Q} \{|A\ra\otimes |C\ra\}\neq 0\,,
\eeq
where $|A\ra$ and $|C\ra$ are the coherent states of the gauge field and for scalar charges respectively. The latter being constructed in canonical formalism as 
\beq
|C\ra=e^{-i\int d^3 x \left( \Phi_{c}(x)\hat{\Pi}(x)-\Pi_{c}(x)\hat{\Phi}(x)+h.c. \right)}|\Omega\ra\,,
\label{globalcohstate}
\eeq
while the former considered to take on the form of \eqref{gaugestate}.

This was a somewhat surprising realization, since the classical charge distribution is expected to source the classical electromagnetic field. And yet it was shown in \cite{Berezhiani:2021zst} that the consistent way of dressing \eqref{globalcohstate} is to replace it with \eqref{dressedC}.

Intuitively, one can understand this by noting that the coherent state \eqref{globalcohstate} is not the state of definite particle number and of $U(1)$ charge. Instead it is the superposition of such states and consequently each and every one of these contributions must be dressed separately. This is precisely what is achieved by \eqref{diracphi} and \eqref{diracpi}.

However, interestingly one can recover the naively expected direct product of two coherent states in a special double-scaling limit
\beq
\label{classlimit}
g\rightarrow 0, \qquad (\Phi_c,\Pi_c)\rightarrow \infty,\qquad g\rho_c=ig\left(\Phi_c\Pi_c-\Phi_c^*\Pi_c^*\right)\rightarrow fixed\,. 
\eeq
Substituting the expanded \eqref{diracphi} and \eqref{diracpi} in powers of $g$ in \eqref{dressedC} and applying the classical limit in question, one arrives at
\beq
|C\ra _g\rightarrow e^{-i\int d^3 x\left( \Phi_{\rm c}\hat{\Pi}-\Pi_{\rm c}\hat{\Phi}+h.c. \right)}\cdot e^{-i\int d^3 x~g\rho_c\left[\frac{1}{\nabla^2}\partial_j\hat{A}_j\right]}|\Omega\ra\,.
\eeq
Furthermore, for localized charge configurations the second exponent can be rearranged into the form of \eqref{gaugestate} by integrating by parts.

This is precisely the direct product of two coherent states we were after, which apparently follows from a consistent, BRST-invariant coherent state albeit merely in the classical limit \eqref{classlimit}.

\section{gauge-fixing and pure-gauge configurations}
\label{gauge_fix}

The point we would like to stress upon here is the consistency with the gauge condition, as the state given by \eqref{gaugestate} is physical for arbitrary $A_j^c$ and $A_0^c$, due to the BRST invariance of $\hat{E}_j$ and $\hat{B}$. In fact, the corresponding configuration could even be a pure gauge mode, with vanishing electric field. But, within the adopted quantisation scheme the following condition must hold
\beq
\partial_0 \la \Psi |\hat{A}_0|\Psi\ra=\partial_j \la \Psi |\hat{A}_j|\Psi\ra\,.
\label{1pointgauge}
\eeq
This condition is dictated by the gauge-fixing terms introduced upon quantization of the theory and it must be obeyed by all physical states, which is why we shall refer to it as \textit{master} gauge. Notice that this equation determines the time-evolution of $\la \Psi |\hat{A}_0|\Psi\ra$. In other words, if we are interested in having a vanishing expectation value for this temporal component at all times, we need to ensure that the vector potential is divergence-less at all times. This is connected to the way we have set up the question, that the specification of the coherent state fixes merely initial conditions for various expectation values. If we are interested in employing the background field method, by decomposing the fields into its expectation value and the fluctuation operator, one can proceed by defining the fluctuation field $\hat{a}_\mu$ via
\beq
\hat{A}_\mu=\bar{A}_\mu(x,t)+\hat{a}_\mu\,, \qquad {\rm where} \qquad \bar{A}_\mu\equiv \la \Psi | \hat{A}_\mu |\Psi\ra\,.
\eeq
In other words, $\bar{A}_\mu$ represents the full 1-point function, instead of being merely a classical background. This decomposition is useful when the state is of high occupancy, with a nonvanishing classical limit for the 1-point function, i.e. $\lim_{\hbar\rightarrow 0}\bar{A}_\mu\neq 0$. 
As a result, the computation of correlation functions can be done iteratively in $\hbar$. Namely, to capture the effects at $\hbar^n$-order, we only need the background field $\bar{A}_\mu$ up to $\hbar^{n-1}$ corrections. If we are, furthermore, interested in having one-point functions to follow a classical solution of a certain gauge (to $\mathcal{O}(\hbar^0)$, of course) then we need to make sure that it is consistent with the gauge condition dynamically. As we already mentioned in the introduction, fundamentally the backgrounds are introduced as quantum states after the quantization and the introduction of the gauge-fixing sector, implying the gauge condition for the background and perturbations to be interconnected. Although, for computing correlation functions of gauge-invariant quantities the gauge condition can be modified after the quantization within the path-integral formalism \cite{Weinberg}. Similar considerations might have important ramifications for quantum effects in cosmological correlation functions \cite{Berezhiani:2024boz}, due to the fact that such computations deal with variables that are gauge-invariant merely to the linear order in perturbations.

The gauge-fixing equation for the 1-point function \eqref{1pointgauge} does not preclude the existence of gauge field configurations that are related by the gauge transformation in the physical Hilbert space. Rather, it informs us on how the expectation value of the temporal component will be evolving. To illustrate the point, let us consider a coherent state of vanishing electromagnetic field 
\beq
|\Psi\ra=e^{-i\int d^3 x\left(A_j^c \hat{E}_j+A_0^c\hat{B}\right)}|\Omega\ra\,,\qquad {\rm with}\qquad A_j^c=\partial_j \alpha(x)\,.
\label{cohA}
\eeq
Thus, the state in question is parameterized by two functions $\alpha$ and $A_0^c$.

Next we would like to demonstrate if and when the pure gauge state in question is physically equivalent to the vacuum. Using integration by parts, the state can be rewritten as
\beq
\label{gaugemode}
|\alpha\ra=e^{-i\int d^3 x\left(-\alpha\partial_j \hat{E}_j+\partial_j(\alpha \hat{E}_j)+A_0^c\hat{B}\right)}|\Omega\ra\,.
\eeq
The first term in the exponent can be expressed in terms of $\hat{B}$ by means of the corresponding Maxwell's equation of the BRST quantization
\beq
\label{Beq}
\partial_0 \hat{B}=i[H,\hat{B}]\,.
\eeq
The part of the Hamiltonian relevant to this equation is readily given by
\beq
\label{HA0}
\hat{H}\subset\int d^3 x\left[\p_j \left(\hat{A}_0\hat{E}_j\right)+\hat{A}_0 \left( -\p_j \hat{E}_j+\hat{J}_0 -\rho_{\rm vac}\right)\right]\,,
\eeq
with $\rho_{\rm vac}$ denoting the vacuum charge counter-term. Notice that the removal of the infinite vacuum charge, which is sometimes removed by the imposition of the so-called normal ordering, requires the introduction of the Lorentz-invariance 'violating' counter-term. However, in reality the violation is merely an illusion as it is not reflected in any physical observable. The necessity of retaining the boundary term in \eqref{HA0} stems from the fact that we are keeping the boundary term in the exponent of \eqref{gaugemode}, i.e. considering $\alpha$ with non-vanishing boundary behaviour.

Therefore, combining \eqref{Beq} and \eqref{HA0} yields
\beq
\label{Bmaxwell}
\partial_j\hat{E}_j=-\partial_0\hat{B}+\hat{J}_0-\rho_{\rm vac} +\int dS^j(z) \hat{E}_j(z)\delta^{(3)}(x-z)\,.
\eeq
Substituting this into \eqref{gaugemode}, the last (surface) term of \eqref{Bmaxwell} cancels the boundary term of  \eqref{gaugemode} for arbitrary $\alpha$, resulting in
\beq
\label{gaugemodesimpl}
|\alpha\ra=e^{-i\int d^3 x\left(\alpha\partial_0\hat{B}+A_0^c\hat{B}\right)}e^{i\int d^3 x\left(\hat{J}_0-\rho_{\rm vac}\right)}|\Omega\ra\,.
\eeq

Taking into account \eqref{bqc}, the conservation of the BRST charge and that $\hat{Q}|\Omega\ra=0$ we can straightforwardly see that
\beq
|\alpha\ra=e^{i\int d^3 x\left(\hat{J}_0-\rho_{\rm vac}\right)}|\Omega\ra+\hat{Q}|\psi\ra\,,
\eeq
where $|\psi\ra$ is some (unphysical) state. The first term, already in the current form, shares some of the properties of the vacuum state. Namely, the expectation value of the gauge field vanishes for it. To bring the similarity a step further, let us notice that for the theory at hand
\beq
\left(\hat{J}_0-\rho_{\rm vac}\right)|\Omega\ra=0\,.
\eeq
For instance, this is straightforward to verify to the leading order in perturbation theory.

Therefore, the coherent state of the pure gauge configuration is physically equivalent to the vacuum, as
\beq
e^{-i\int d^3 x\left((\partial_j\alpha) \hat{E}_j+A_0^c\hat{B}\right)}|\Omega\ra=|\Omega\ra+\hat{Q}|\psi\ra\,,
\eeq
with $|\psi\ra$, once again, being some state. As a result, the nilpotence of the BRST charge $\hat{Q}^2=1$ ensures that the $S$-matrix elements involving $|\alpha\ra$ are identical to the ones with $|\Omega\ra$. This statement is obviously applicable to non-vacuum states as well. Namely, for any physical (i.e. BRST-invariant) state $|phys\ra$, we can contract a state that differs from it by a pure gauge configuration analogously 
\beq
|phys\ra_\alpha=e^{-i\int d^3 x\left( (\partial_j \alpha)\hat{E}_j+A_0^c\hat{B}\right)}|phys\ra\,.
\eeq
For the same reasons, the $S$-matrix elements involving such states are independent of $\alpha$. The expectation values of the vector potential, as well as any other correlation function of gauge non-invariant operators, on the other hand are explicitly dependent on $\alpha$.

\section{Topological Defects}
\label{top_def}

Having discussed some of the essential aspects of the coherent state description of classical configurations, we would like to focus our attention on topological defects. Classically, they are field configurations found as solutions to classical equations of motion. Fundamentally, we need to associate quantum states with them. 

The corpuscular description of topological solitons as $N$-particle states with $N$ given by the inverse coupling was initiated in  
\cite{Dvali:2011aa}. The systematic construction \cite{Dvali:2015jxa} of corresponding coherent states shows that the inverse coupling indeed sets the number $N$ of quanta contributing to the energy of the soliton. These have the wavelengths given by the size of the soliton
$L$, so that the mass of the soliton is $M \sim  N/L$. 
However, at the same time, the topological charge is determined by the 
deep infrared quanta with infinite occupation number.  
   In quantum theory, the localized soliton can be treated 
 as an approximate eigenstate of the Hamiltonian with the precision 
 $1/N$, since the spread-out time is $t = L^2M \sim N L$. 
 
 Therefore, it is natural to attempt to describe solitons by means of 
 coherent states due to possession of classical features.
We would like to begin with a construction of the coherent state description of Abelian global strings. For this we consider the scalar field potential that facilitates spontaneous breaking of the $U(1)$ symmetry
\beq
V=\frac{\lambda}{2}\left(|\hat{\Phi}|^2-v^2 \right)^2\,.
\eeq
The theory with such a potential has degenerate vacua $|\alpha\ra$ satisfying 
\beq
\la \alpha |\hat{\Phi} |\alpha \ra=ve^{i\alpha}\,, \qquad {\rm with} \qquad \alpha\in[0,2\pi)\,.
\label{alphavacua}
\eeq
These states can be introduced as a set of orthonormal, degenerate, eigenstates of the Hamiltonian. In fact, for perturbative calculations we only need to introduce one of these vacua explicitly. 

However, as topological defects represent states that wind around vacuum manifold we need to identify the way for interpolating between these vacuum states. Generically, the charge operator of the spontaneously broken symmetry generates the excursion between degenerate vacua as
\beq
|\alpha\ra=e^{-i\alpha\int d^3 x ~i(\hat{\Phi}\hat{\Pi}-\hat{\Pi}^\dagger\hat{\Phi}^\dagger)}|0\ra\,.
\label{Qvacua}
\eeq
The expectation value of the Hamiltonian (or any $U(1)$ invariant operator) is degenerate in $\alpha$-parameter and the relation \eqref{alphavacua} is a direct consequence of the Baker-Campbell-Hausdorff formula.

\subsection*{An alternative approach}

Another possible representation of these states could be via the field-displacement operator. In other words, we could think of one vacuum state as a coherent state constructed around the other. In the infinite volume, such coherent states are orthogonal to each other, because they differ by the infinite number of zero-energy particles and the overlap of such coherent states vanishes. The reason one might be inclined to approach it this way, as proposed in \cite{Dvali:2015jxa} for the degenerate discreet vacua, is the desire to construct topological defects such as domain walls as coherent states.

In our case of connected vacuum manifold, in the limit of decoupled gauge field $g\rightarrow 0$ we could construct the vacuum states around the $\alpha=0$ vacuum as
\beq
|\alpha\ra = e^{-i\int{\rm d}^3x\left(\Phi_c\hat{\Pi}+\Phi_c^*\hat{\Pi}^\dagger\right)}|0\ra\,, \qquad {\rm where} \qquad \Phi_c=v\left( e^{i\alpha}-1\right)\,,
\label{vacua}
\eeq
and the vacuum state $|0\ra$ satisfies $\la 0| \hat{\Phi} |0 \ra=v$. It is straightforward to see that the expectation value of the potential in $|0\ra$ diverges, due to the appearance of the correlation functions at space-time coincidence, which is readily absorbed by the renormalization of the vacuum energy.

It is important to keep in mind that we cannot readjust it to accommodate other vacua. In fact, ideally all $\alpha$-states defined by \eqref{vacua} should have finite energy once we renormalize the ground state energy for $|0\ra$. Unfortunately they do not, and we have to tweak the states, presumably generalizing \eqref{vacua} to squeezed coherent states.

\subsection*{Moving on using charge operator construction}

Instead of attempting to go down the aforementioned squeezed road, we proceed with the use of the charge operator to generate different vacua\footnote{However, one must be fully aware that in case of spontaneously broken discrete symmetries this is not an option and we have to follow the path of constructing squeezed states. In principle, this is not an issue, especially since the need for squeezed states is unavoidable within interacting field theories by consistency \cite{Berezhiani:2020pbv,Berezhiani:2021gph,Berezhiani:2023uwt}.}. One can even generalize the constant parameter $\alpha$ in \eqref{Qvacua} to the function of spacial coordinates
\beq
|\alpha\ra \rightarrow e^{-i\int d^3 x ~i\alpha (x)(\hat{\Phi}\hat{\Pi}-\hat{\Pi}^\dagger\hat{\Phi}^\dagger)}|0\ra\,.
\label{Qvacua1}
\eeq
In this state, the expectation value of the field becomes
\beq
\la \hat{\Phi}(x) \ra=e^{i\alpha(x)}v\,.
\eeq
Now we can go ahead with the construction of topological defects as coherent states.

\subsection*{Global string in $g\rightarrow 0$ limit}

Starting off in $g\rightarrow 0$ limit, we can construct a state of the global string as
\beq
|S\ra = e^{-i\int d^3 x ~i\alpha (x)(\hat{\Phi}\hat{\Pi}-\hat{\Pi}^\dagger\hat{\Phi}^\dagger)} e^{-i\int d^3 x f(x)(\hat{\Pi}+\hat{\Pi}^\dagger)} |0\ra\,.
\label{globalstring1}
\eeq
The expectation value in this state reduces to
\beq
\la S | \hat{\Phi}(x) | S \ra=e^{i\alpha(x)}\la 0 | \hat{\Phi}(x)+f(x) | 0 \ra=e^{i\alpha(x)} (v+f(x))\,.
\eeq
Let us note that the order of exponentials in \eqref{globalstring1} is important. If we were to reverse the order, additional operators would make appearance.

Now the string configuration would correspond to the choice of $\alpha(x)$, that winds around the vacuum manifold and $f(x)$ given by the solution to the equations of motion with appropriate limiting behavior in the center of the soliton and at the boundary. To be specific, a straight string configuration (minimizing the energy density) in cylindrical coordinates $(\rho,\theta,z)$ can be characterized by mapping of the vacuum manifold onto the spacial boundary with winding number $n$, by means of $\alpha(x)=n\theta$ and $f$ being a function of $\rho$ only. The detailed form of $f(\rho)$ follows from solving classical equation of motion with adequate boundary conditions. Being a textbook material, we will not reiterate further details here.

As we know the total energy of the global string, or a vortex if we were to consider the theory in $(2+1)$-dimensions, receives a logarithmically diverging contribution from large distances from the soliton. In this regard the state \eqref{globalstring1} is no different. The resulting logarithmic divergence can be regulated in the phenomenologically interesting setting by the introduction of the network of such strings and anti-strings. In other words, they could be formed in the cosmological setting.

\subsection*{Global string for $g\neq 0$}

Next we reintroduce the gauge field and see how the construction changes. Classically the global strings would be unchanged, but there will emerge new string solutions that have nonvanishing magnetic field; known as Nielsen-Olesen strings \cite{Nielsen:1973cs}. In quantum theory, on the other hand, the introduction of the gauge field introduces a constraint that physical states must satisfy. In BRST quantization, as we have already discussed, the physical states are to be annihilated by the BRST charge $\hat{Q}$. As we have already done above, we achieve this by replacing the matter operators by their gauge-invariant counterparts. The charge density is gauge-invariant and thus there is no need to modify the first exponential factor of \eqref{globalstring1}. The second factor on the other hand needs to be made gauge-invariant, which can be achieved using Dirac operators as before. As a result the BRST-invariant state corresponding to the global string can be cast in the form
\beq
|S\ra_{\rm global} = e^{-i\int d^3 x ~i\alpha (x)(\hat{\Phi}\hat{\Pi}-\hat{\Pi}^\dagger\hat{\Phi}^\dagger)} e^{-i\int d^3 x f(x)(\hat{\Pi}_{\rm g}+\hat{\Pi}_{\rm g}^\dagger)} |0\ra\,.
\label{globalstring2}
\eeq
One might expect that the expectation value of the Hamiltonian in this state should be the same as in the absence of the gauge field. In other words, we might expect that it reduce to the classical energy after renormalizing the divergencies by the adjustment of counterterms.

To verify this, we can go ahead and compute the expectation value of $\hat{H}$. The only tricky terms of \eqref{hamilton} that require extra care are the ones that depend on the electric field operator, due to the fact that $\hat{\Pi}_{\rm g}$ explicitly depends on $\hat{A}_j$. These terms result in what seem to be (unrenormalizable) singularities discussed in \cite{Berezhiani:2023uwt}, which can be remedied by the squeezing and other non-Gaussian modifications of the coherent state. In this work, we simply assume that such adjustments are possible.

\subsection*{Nielsen-Olesen string}

This brings us to the construction of the fully fledged gauge string, as a straightforward extension of \eqref{globalstring2} by introducing the application of the electromagnetic field displacement operator that posits the vector field configuration (magnetic field) that alleviates the aforementioned logarithmic divergence of the energy of the global string. Namely, we can parameterize the desired state as
\beq
|S\ra_{\rm NO} = e^{-i\int d^3 x\left(A_j^c \hat{E}_j\right)}e^{-i\int d^3 x ~i\alpha (x)(\hat{\Phi}\hat{\Pi}-\hat{\Pi}^\dagger\hat{\Phi}^\dagger)} e^{-i\int d^3 x f(x)(\hat{\Pi}_{\rm g}+\hat{\Pi}_{\rm g}^\dagger)} |0\ra\,,
\label{globalstring3}
\eeq
where $f(x)$ is a c-number function that corresponds to the axisymmetric modulus of the scalar field configuration of the Nielsen-Olesen vortex configuration, $\alpha(x)$ is the mapping of the vacuum manifold onto the spatial boundary (with the simplest mapping being the polar angle) and $A_j^c$ is the vector field configuration of the $U(1)$ string.

\section{Understanding Topological Charge via Occupation Number} 
\label{occup_numb}

 As suggested in  \cite{Dvali:2015jxa}, in coherent state description of solitons, the topological charge can be understood as  a singularity in occupation number of certain zero-momentum modes. 
  We wish to incorporate this view in the present picture.
We shall also show how to reconcile the divergence in the occupation number with the finiteness of energy of a topological defect in a theory with a mass gap, as this is the case with the Nielsen-Olesen vortex. 
In this case, the key is in the  gauge redundancy due 
 to which the topological charge in an asymptotic
 vacuum is realized by a locally-pure-gauge winding 
 configuration. 
  In order to see this, let us first understand the coherent 
  state picture of the vacuum degeneracy in the language of 
  a Goldstone field.  We start with the case of a global $U(1)$. 

The coherent state picture of the vacuum can 
be understood, following the steps of \cite{Dvali:2015rea}, in terms of the Goldstone field. Namely, the spontaneously broken $U(1)$ symmetry is realized as the shift symmetry of this gapless degree of freedom $\varphi$ introduced by means of the following decomposition
\beq
\Phi=\frac{1}{\sqrt{2}}(v+h)e^{i\varphi/v}\,,
\eeq
with $h$ denoting the massive Higgs boson. The Noether charge, for the symmetry in question, is related to the conjugate momentum $\hat{\Pi}_\varphi$ of the Goldstone mode as follows
\beq
\hat{Q}=i\int d^3 x (\hat{\Phi}\hat{\Pi}-\hat{\Pi}^\dagger\hat{\Phi}^\dagger)=v\int d^3x~ \hat{\Pi}_\varphi\,.
\label{Q-lin}
\eeq
For the occupation number analysis, we utilize the Schr\"odinger picture mode decomposition
\beq
\hat{\Pi}_\varphi(\vec{x})=\int \frac{d^3k}{(2\pi)^3}(-i)\sqrt{\frac{k}{2}}\left( \hat{a}_{\vec{k}}e^{i\vec{k}\cdot \vec{x}}-\hat{a}^\dagger_{\vec{k}}e^{-i\vec{k}\cdot \vec{x}} \right)\,,\qquad {\rm with} \qquad [a_{\vec{k}},a^\dagger_{\vec{k}'}]=(2\pi)^3\delta^{(3)}(\vec{k}-\vec{k}')\,.
\eeq
As a result, due to the invariance of the Hamiltonian under the spontaneously broken $U(1)$, we arrive at the following exact expression for the charge operator in terms of the creation-annihilation operators of the Goldstone mode
\beq
\hat{Q}=-\frac{iv}{\sqrt{2}}\lim_{\vec{k}\rightarrow 0}\sqrt{|\vec{k}|}\left(\hat{a}_{\vec{k}}-\hat{a}^\dagger_{\vec{k}} \right)\,.
\eeq
Identifying a vacuum state $|0\ra$ as the state with vanishing occupation number of Goldstone bosons, we have
\beq
\hat{Q}|0\ra=\frac{iv}{\sqrt{2}}\lim_{\vec{k}\rightarrow 0}\left(\sqrt{|\vec{k}|}\hat{a}^\dagger_{\vec{k}}\right)|0\ra\,.
\eeq
In other words, the charge operator creates a single zero-momentum Goldstone mode. Naively, one might think that the above expression vanishes due to the appearance of $\sqrt{|\vec{k}|}$. However, this factor is part of the Lorentz invariant normalization of single-particle states $\la\vec{k}|\vec{k}'\ra=2 k(2\pi)^3\delta^{(3)}(\vec{k}-\vec{k}')$.

    Once we have identified one of vacuum state, in this case $|0\ra$, we can generate the rest of the vacuum manifold enumerated by the parameter $\alpha$ as follows
 \beq
|\alpha \ra = e^{-i \alpha \hat{Q}} |0\ra\,.
\label{Cstate}
\eeq
This is precisely what we did in the previous section. However, the Goldstone field representation \eqref{Q-lin}, manifests the coherent state nature of the vacuum manifold \eqref{Cstate}, built over the vacuum of Goldstone bosons.

 In order to compute the overlap of two states with different values of $\alpha$ and $\alpha'$, one needs to be cautious with zero-momentum limit. 
 We arrive 
\beq
\la \alpha' |\alpha\ra =\lim_{\vec{k}\rightarrow 0} e^{- \frac{v^2}{4}k(2\pi)^3\delta^{(3)}(0) (\alpha - \alpha')^2}\,,
\label{product}
\eeq
where $\delta^{(3)}(0)$ is the momentum space delta-function at the singularity and is the measure of the infinite spatial volume. As a result, the exponential factor is nonzero only for $\alpha=\alpha'$, in which case the whole expression becomes equal to the unity.

 In order to make the above-discussed infrared regularization absolutely transparent, we  giving the Goldstone boson a small mass $m$ and consider the system in compact two-dimensional space of size $R$. At the end of the day, we shall remove the regulators by taking $m$ to zero and $R$ to infinity. 
Of course the mass term has to be understood as the first non-trivial term in the expansion  
of a periodic potential, such as $-\cos(\phi/v)$. 
In the full theory this potential comes from 
a small explicit breaking of $U(1)$-symmetry which we shall later set to zero. 
 In $m\rightarrow0$ limit, the 
$U(1)$ symmetry is realized as the shift symmetry
$\varphi \rightarrow \varphi + const$. 
The corresponding Noether current is obeys $\partial^{\mu}J_{\mu} \, = \,- 
m^2\varphi$. 
 The Schr\"odinger picture mode decomposition of
 $\varphi$, in the regulated case, becomes
 \beq
\hat{\varphi} = \int 
\frac{d^2(\vec{k}R)}{2\pi R\sqrt{2\omega_k}} 
 \left(e^{i \vec{k}\vec{x}} \hat{a}_{\vec{k}}
 + h.c. \right),
\label{modes}
\eeq
where $\omega_{\vec{k}} = \sqrt{m^2 + |\vec{k}|^2}$, 
$R$ is the radius of the periodic space and creation-annihilation operators have been normalized to be dimensionless. The charge takes the following 
form 
   \beq
\hat{Q} = \int d^2 x J_0 = 
\int d^2 x \Pi_\varphi = 
-i 2\pi R \sqrt{\frac{m}{2}} \left(\hat{a}_{0} -\hat{a}_{0}^{\dagger}  \right)\,,
\label{Qphi}
\eeq
where $\Pi_\varphi$ is the canonical conjugate momentum of the Goldstone field.
Now, this operator can be used to create shifted coherent state \eqref{Cstate}, albeit in the finite volume and with a mass regulator in palce.

 In this case, the state constructed as \eqref{Cstate} represents a coherent state with 
 mean occupation number of zero momentum quanta given by  $n_0 = (2\pi R)^2 m \alpha^2$.  
 In general for two states corresponding to two different values $\alpha $ and $\alpha'$ have non-trivial overlap.
 
 The two states become strictly orthogonal in the infinite volume limit   $R^2 m \rightarrow  \infty$.
 Moreover, taking additionally $m  \rightarrow 0$, they become the degenerate vacua.  
 
 Thus, the vacua of Goldstone theory 
 can be viewed as coherent states with infinite relative 
 occupation numbers of the zero-momentum modes 
 of the Goldstone field. 
  
  Notice that in any given vacuum of topologically-trivial sector,  it is always possible to normalize the occupation number to zero by an appropriate shift of the field.  
   However, this is not possible in a topologically-non-trivial sector.   For example, in the presence of a vortex
  with unit topological charge,  due to non-zero winding number of the Goldstone field, the asymptotic vacua
  necessarily differ by the occupation numbers of the Goldstone mode and this difference cannot be removed 
  by any non-singular field-redefinition.   
  Correspondingly,  the vortex considered as a coherent state,  necessarily involves an infinite occupation number of the zero-momentum Goldstone modes.
     
     This story persists in the Higgs theory. There too, 
   the  topological charge translates into infinite occupation number of zero-momentum modes. 
     Despite the fact that the theory has a mass gap, 
     the infinite occupation number does not result into
     an infinite energy of the vortex, in $2+1$-dimensions or per unit length in $3+1$-dimensions.  The reason is the  
  gauge redundancy.  Due to it, 
  the field deformations  $\varphi \rightarrow \varphi 
  + \kappa(x)$  that in globally-symmetric $U(1)$  theory would
  describe the local excursions into neighbouring Goldstone 
  vacua, in the Higgs theory describe the pure-gauge transformations that locally cost no energy. 
  However, the vacuum cannot be deformed into a 
 trivial one due to the topological obstruction. 
  As a result, the singularity in the occupation number of zero-momentum modes is unremovable. 
  
\section{Generalization to Instantons} 
\label{instant}

Our construction can be used for obtaining 
  the BRST-invariant 
 consistent corpuscular resolution of 
 an instanton. 
  The concept of corpuscular resolution 
  of instantons was introduced in \cite{Dvali:2019ulr}. 
 This resolution acquires a well defined meaning 
 when the two facts are superimposed.  
  First, is the understanding of an instanton 
as of a passing-though soliton in theory with one
additional space dimension \cite{Dvali:2007nm}. 
The second point is the corpuscular 
     resolution of this soliton. 

Let us consider the Higgs model in 
$1+1$-dimensions. 
The vacuum of the theory 
admits topologically non-trivial pure-gauge
winding configurations, 
\beq \label{windingN}
\Phi=\frac{1}{\sqrt{2}}\,v \,e^{i\varphi(x)/v}\,,~ 
A_{\mu} = \frac{1}{g} \partial_{\mu} \frac{\varphi(x)}{v}\,, 
\eeq
with 
$\int dx A_{x}  = \frac{1}{g v} \int dx \partial_{x} \varphi = 2\pi n/g$,
where $n$ is an integer.   Notice that since the Goldstone 
field $\varphi/v$ is defined modulo $2\pi$, the asymptotic 
vacua are physically equivalent. The size of the region over which the phase changes can be shrunk arbitrarily by a non-singular gauge transformation.   For changing the winding number 
by a smooth deformation of the field, it suffices that 
the modulus of the field $\Phi$ touches zero at 
one point.  This implies that the vacua with different 
$n$ are separated by a finite energy barrier and can
be changed by tunneling.  The corresponding instanton has 
a finite Euclidean action. 

This is very different from the case of the winding number 
of the vortex in $2+1$-dimensions. There the topological charge is 
conserved also at the quantum level. This can be understood 
in two ways. First, the change of the winding number by a smooth deformation would require vanishing of the modulus across a 
line that extends from the origin ($\rho=0$) to infinity ($\rho = \infty$).  
 Such a deformation obviously costs an infinite energy which 
 results in an infinite barrier separating the sectors with different $n$.

 An alternative language for understanding the difference is through the coherent state resolution.  Unlike the vortex in $2+1$-dimensions, the pure-gauge vacuum configuration of $1+1$-dimensional theory  
can be resolved as a coherent state with finite occupation number. 
 This distinction is clear also from the fact that a vortex corresponds to 
 collection of one-dimensional windings around an infinite number 
 of concentric circles.  
 
  The winding number of $1+1$-dimensional theory 
  $n$ is changed  by transitions 
  mediated by instantons.  The 
  instantons are the Nielsen-Olesen vortices in two-dimensional 
  Euclidean space. 
    Thus, the instantons of $1+1$-dimensional theory  
    are in one-to-one correspondence with 
 solitons (vortices) of $2+1$ dimensional theory. 
 This mapping is more than just a coincidence between the solutions.
  The point is that the winding number changing trajectories  mediated by instantons in a lower dimensional theory can be 
  obtained with the actual passage of high-dimensional  
  solitons.  
    If we identify the $1+1$-dimensional Higgs theory 
    as a world-sheet theory of a line embedded in $2+1$-dimensions, 
    the winding number can be changed by moving a high-dimensional vortex across it. 

     Notice that the higher dimensional theory is just an auxiliary 
     concept for visualizing the process of the change in winding number.  In particular, it makes it clear why the changes in the 
     winding number cost a finite action of Euclidean solution. 
      The reason is that from the higher dimensional perspective 
      the initial and final configurations belong to the same topological sector since the number of vortices in 
      high-dimensional space is unchanged. The cost of the action 
      for the process solely comes from a finite displacement 
      of the vortices in extra space.

    The above mapping allows to generalize the concept of 
     corpuscular resolution of solitons to the similar resolution 
     of instantons \cite{Dvali:2019ulr}. Of course, the constituents of the instantons 
     are processes rather than particles but this does not change 
     the essence of the collective effect. 
     Each constituent of an instanton accounts for a tunneling 
     of a corresponding soliton constituent across the 
     lower dimensional space. The instanton describes a collective tunneling. 

       Of course, to the leading order, the transition rate is reliably computable within the ordinary 
       instanton calculus of $1+1$-dimensional theory, without 
       invoking additional dimensions. 
     However, the corpuscular resolution of the instanton 
     via a BRST-invariant coherent state and its uplifting 
     to a high-dimensional vortex allows the visualization of 
     the topology of the process. 

     For example, assuming that the space coordinate of the $1+1$-dimensional theory is compactified 
     on a circle,  from the point of view of a $1+1$-dimensional observer, the winding number counts the number of
  $2+1$-dimensional vortices 
     inside the circle. Correspondingly, the change of the winding number is due to a migration of a vortex across the circle. 
  Such a migration does not change the  asymptotic  occupation number of constituents  of the vortex in $2+1$-dimensional theory. 
 Thereby, the corresponding transition has a finite rate. 

\section{Outlook}
\label{outlook}

In this work we have revisited coherent states within the BRST quantization of electromagnetism. In our framework the theory is quantized around the vacuum once and for all, with the gauge-fixing condition that is convenient for a particular computation.
 We referred to this as a \textit{master} gauge condition.
The classical backgrounds are then introduced as BRST-invariant coherent states built over the vacuum. Correspondingly, the 
background fields correspond to one-point expectation values of the field-operators over such states. Moreover, the \textit{master} gauge-fixing condition, introduced upon quantization around the vacuum, will be obeyed by one-point functions at all times. This implies that, if we are interested in recovering the background field in a particular gauge, we need to start with appropriate \textit{master} gauge-fixing term. 

In other words, the evolution of the system including 
its departures from classicality are encoded in the quantum 
evolution of the BRST-invariant state. This fully quantum picture, 
in certain approximation should capture the physics of standard 
semi-classical treatment in which the system is split in classical background and the perturbations quantized on top of it.

At the tree-level our treatment is expected to  agree with the standard procedure. However, the differences coming from the 
resolution of classical background as of BRST-invariant  coherent state 
over a master gauge-fixed quantized field can potentially set in at higher orders. 

 As part of understanding of the recovery of the classical behavior, 
we have also found a BRST-invariant coherent state description for pure-gauge configurations in QED. Such configurations are interesting for 
number of reasons.  In particular, they serve as proxies for 
understanding the coordinate reparametrization of General Relativity \cite{Berezhiani:2024boz}. 

 At the same time, such pure-gauge coherent states play an important role in understanding 
the topological structure of vacuum in BRST-invariant description.  
 As an example, we constructed a $U(1)$-vortex state as a BRST-invariant coherent state. To be precise, the modulus of the corresponding Higgs field is posited by the standard field-displacement operator, while the winding is generated by the charge density operator of the spontaneously broken symmetry. This is facilitated by the fact that the charge of the broken symmetry generates the excursion in the vacuum manifold. 

This BRST-invariant coherent state in vacuum approaches a topologically non-trivial pure-gauge configuration with non-zero winding number. 
  This description makes it very explicit that
  the topological charge translates as an infinite occupation number 
  of infinite-wavelength quanta in the coherent state describing the soliton \cite{Dvali:2015jxa}. 

 This origin of topological charge also sheds light on 
 the instanton transitions between would-be vacua with different topological numbers in gauge theories. The finiteness of the rate
 is due to finite differences of occupation numbers in coherent states 
 describing these winding configurations. This amounts to the finiteness of the instanton action which can be mapped on a 
 a soliton process in one higher dimension 
 \cite{Dvali:2007nm, Dvali:2019ulr}. 
   
It is important to keep in mind that the situation is more subtle in the disconnected vacuum manifold, i.e. in the case of the spontaneously broken discreet symmetry. In that case one should consider relating different vacua by means of the field displacement operator, in order to be able to construct the corresponding topological defects, domain walls, as coherent states \cite{Dvali:2015jxa}. However, as we have argued, the field-displacement must be supplemented with the nontrivial squeezing, when viewing one vacuum as a coherent state constructed on top of the other.

It must be also stressed that the background dependent squeezing and non-Gaussian modifications of the constructed coherent states, along the lines of \cite{Berezhiani:2023uwt}, are essential for perturbative renormalizability.

\section*{Acknowledgements}

We would like to thank Giordano Cintia, Giacomo Contri and Archil Kobakhidze for stimulating and fruitful discussions. O.S. would like to thank the Arnold Sommerfeld Center at LMU and the Max-Planck-Institute for Physics, Munich for the hospitality extended during the completion of this work. This work was supported in part by the Humboldt Foundation under the Humboldt Professorship Award, by the European Research Council Gravities Horizon Grant AO number: 850 173-6, by the Deutsche Forschungsgemeinschaft (DFG, German Research Foundation) under Germany’s Excellence Strategy - EXC-2111 - 390814868, Germany’s Excellence Strategy under Excellence Cluster Origins  EXC 2094 – 390783311 and 
the Australian Research Council under the Discovery Projects grants DP210101636 and DP220101721. \\

 Disclaimer: Funded by the European Union. Views
and opinions expressed are however those of the authors
only and do not necessarily reflect those of the European Union or European Research Council. Neither the
European Union nor the granting authority can be held
responsible for them.

\end{document}